\documentclass[a4paper,twocolumn]{article}

\usepackage{jsaiac}

\usepackage{color}
\usepackage{graphicx}
\usepackage{algorithm}
\usepackage{algorithmic}
\usepackage{booktabs}
\usepackage{multirow}
\usepackage{subcaption}
	
\title{Automated Web Application Testing: End-to-End Test Case Generation \\with Large Language Models and Screen Transition Graphs}

\address{Minh Le Nguyen, JAIST, Nomi, 07651211, nguyenml@jaist.ac.jp}

\author{%
Nguyen-Khang Le\first
\and
Quan Minh Bui \second
\and
Minh Ngoc Nguyen\first
\and
Hiep Nguyen\first
\and
Trung Vo\first
\and
Son T. Luu\first
\and
Shoshin Nomura \second
\and
Minh Le Nguyen \first
}

\affiliate{
\first{} Japan Advanced Institute of Science and Technology
\and
\second{} Amifiable Inc.
}

\begin{abstract}
Web applications are critical to modern software ecosystems, yet ensuring their reliability remains challenging due to the complexity and dynamic nature of web interfaces. Recent advances in large language models (LLMs) have shown promise in automating complex tasks, but limitations persist in handling dynamic navigation flows and complex form interactions. This paper presents an automated system for generating test cases for two key aspects of web application testing: site navigation and form filling. For site navigation, the system employs screen transition graphs and LLMs to model navigation flows and generate test scenarios. For form filling, it uses state graphs to handle conditional forms and automates Selenium script generation. Key contributions include: (1) a novel integration of graph structures and LLMs for site navigation testing, (2) a state graph-based approach for automating form-filling test cases, and (3) a comprehensive dataset for evaluating form-interaction testing. Experimental results demonstrate the system's effectiveness in improving test coverage and robustness, advancing the state of web application testing.
\end{abstract}

\def\BibTeX{{\rm B\kern-.05em{\sc i\kern-.025em b}\kern-.08em%
 T\kern-.1667em\lower.7ex\hbox{E}\kern-.125emX}}
\def\JBibTeX{\leavevmode\lower .6ex\hbox{J}\kern-0.15em\BibTeX}
\def\LaTeXe{\LaTeX\kern.15em2$_{\textstyle\varepsilon}$}

\begin{document}
\maketitle

\section{Introduction}

Web applications are integral to modern software ecosystems, powering diverse services ranging from e-commerce to social networking. Ensuring their reliability and functionality is paramount, as users expect seamless operation under varying conditions. However, the complexity and dynamic nature of web interfaces make manual testing a significant bottleneck. This has spurred the need for scalable and automated testing solutions, driving research into novel methodologies for web application testing.

Traditional testing approaches, such as static and dynamic analysis techniques, have been foundational in identifying vulnerabilities and ensuring functionality. Early work \cite{ricca2001} emphasized the importance of comprehensive analysis techniques, while subsequent studies \cite{andrews2005} introduced systematic test generation methods to improve coverage. More recent advancements have focused on automated frameworks that leverage pre-recorded test cases to reduce human intervention \cite{smith2022}. Despite these efforts, traditional methods struggle to scale to large, interactive web applications, often resulting in gaps in test coverage and undetected vulnerabilities. Existing automated tools, such as Selenium, enable programmatic interaction with web elements but typically require manually generated test cases. This process is time-consuming, error-prone, and ill-suited for dynamic web environments. To address these limitations, researchers have explored generative AI techniques for test case creation. For instance, \cite{lee2022} proposed an adaptive approach where AI models dynamically generate test cases in response to application changes, improving coverage and robustness. However, significant challenges remain, particularly in handling dynamic navigation flows and complex form interactions.

\begin{figure*}[ht]
    \centering
    \includegraphics[width=1\linewidth]{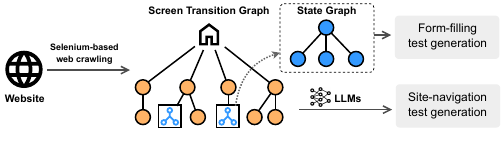}
    \caption{Overview of the test generation process, focusing on two types of tests: Navigation tests and form action tests.}
    \label{fig:overview}
\end{figure*}

The advent of large language models (LLMs), such as ChatGPT and GPT-4 \cite{openai2023gpt4}, has opened new avenues for automating complex tasks. Researchers have developed LLM-based autonomous agents \cite{autogpt2022} capable of executing intricate workflows \cite{qin2023toollm, schick2023toolformer}. In the context of web testing, recent studies have explored text-based web browsing environments and instructed LLM agents to perform web navigation \cite{nakano2021webgpt, gur2023realworld, zhou2023webarena, lu2023chameleon}. A key challenge in these works is managing complex and verbose HTML structures, with solutions including HTML simplification and structuring \cite{nakano2021webgpt, zhou2023webarena, gur2023realworld, deng2023mind2web}. Notably, \cite{he-etal-2024-webvoyager} introduced WebVoyager, an end-to-end automated web testing framework employing multimodal LLMs to interact with real-world websites.

Despite these advancements, existing methods face critical limitations. Many approaches struggle to model dynamic navigation flows effectively, leading to incomplete coverage of critical user pathways. Additionally, handling complex form interactions—such as conditional fields that appear based on prior inputs or mandatory fields requiring specific validation logic—remains a significant challenge. Furthermore, current systems often treat navigation modeling and form validation as separate processes, lacking a unified workflow to address these interconnected aspects of web testing.

Two of the most common and critical actions in web applications are \textit{site navigation} (moving between pages) and \textit{form filling} (interacting with input fields). These actions are fundamental to user experience and functionality, yet they pose unique challenges for automated testing. This paper addresses these challenges by presenting an automated system designed to generate test cases for site navigation and form filling. For site navigation, the system employs \textit{screen transition graphs} to model navigation flows between web pages, enabling the generation of comprehensive test cases. For form filling, the system integrates a \textit{state graph} to handle complex form-filling scenarios and automates the generation of Selenium scripts. Together, these components provide a scalable and efficient solution for web application testing. The key contributions of this work are as follows:

\begin{itemize}
    \item A novel approach leveraging graph structures to model \textit{screen transition graphs}, combined with large language models (LLMs) for automated test scenario generation in site navigation.
    
    \item A novel methodology utilizing \textit{state graphs} to model potential state changes in complex conditional forms, integrated with LLMs to generate high-quality Selenium test cases.
    
    \item The development of a comprehensive dataset for evaluating form-interaction testing, facilitating robust benchmarking and validation of the proposed system.
\end{itemize}

\section{Related Work}

Testing web applications is a critical yet time-intensive task in software development. Over the years, various approaches have been explored to streamline this process. Early research in web application testing focused on static and dynamic analysis techniques to identify vulnerabilities and ensure functionality. For instance, \cite{ricca2001} emphasized the importance of comprehensive analysis techniques for testing web systems, while \cite{andrews2005} introduced systematic test generation methods aimed at improving test coverage.

Recent advancements have introduced innovative methods for automating web application testing. Tools like Selenium have been pivotal in enabling programmatic interaction with web elements. However, these tools often require manually generated test cases, which are time-consuming and error-prone. To address this, \cite{marchetto2012} provided a comprehensive survey highlighting the challenges and potential solutions for web application testing, setting the stage for subsequent research. More recently, researchers have proposed automated frameworks driven by pre-recorded test cases, significantly reducing human intervention. For instance, \cite{smith2022} demonstrated a framework that utilizes recorded user interactions to automatically generate reusable test scripts. This approach not only reduces manual effort but also enhances test efficiency.

Another promising avenue involves leveraging generative AI for automated test case creation. \cite{lee2022} presented a methodology where AI models adaptively generate test cases in response to changes in web applications. This dynamic approach improves testing comprehensiveness and ensures robustness against application updates. Additionally, \cite{he-etal-2024-webvoyager} introduced WebVoyager, an end-to-end automated web testing framework that employs multimodal large language models (LLMs) as agents to interact with real-world websites. Despite these advancements, significant limitations remain in existing methods. First, many approaches struggle to effectively handle dynamic navigation flows, which are a common feature of modern web applications. This often results in incomplete coverage of critical user pathways. Second, existing solutions frequently fail to address the complexities of advanced form interactions, such as conditional fields that appear based on prior inputs or mandatory fields requiring specific validation logic. These limitations reduce the robustness and reliability of automated testing frameworks. Third, most current systems lack integration between navigation modeling and form validation, treating these as separate processes rather than a unified workflow.

Our work addresses these gaps by introducing a unified system that integrates \textit{screen transition graphs} for navigation modeling and \textit{state graphs} for robust handling of advanced form-filling scenarios. By automating the generation of Selenium scripts and ensuring comprehensive test coverage, our approach provides a scalable and efficient solution tailored to the challenges of modern web application testing. This integration not only fills critical gaps in existing methodologies but also enhances the reliability and scalability of automated testing processes.

\section{Automated Test Generation}

Figure \ref{fig:overview} illustrates the overview of the test generation process, which involves two primary types of test cases: \textit{site navigation test cases} and \textit{form-filling test cases}. For a given website, the process begins with the construction of a \textit{screen transition graph} through Selenium-based web crawling. In this graph, each node represents a web page and encapsulates metadata such as the URL, HTML content, and other relevant information. Edges between nodes denote navigational paths, such as links in anchor tags or buttons, that enable users to transition from one page to another. When a web page contains a form, it is associated with a \textit{state graph}, a sub-graph that models the states and interactions of the form. In the state graph, nodes represent distinct form states, while edges represent user interactions that transition the form from one state to another. The screen transition graph is utilized to generate navigation test cases, while the state graph is employed to generate form action test cases. This dual-graph approach ensures comprehensive coverage of both navigation flows and form interactions, providing a robust foundation for automated web application testing.


\subsection{Site-navigation test generation}

\begin{figure}[ht]
    \centering
    \includegraphics[width=1\linewidth]{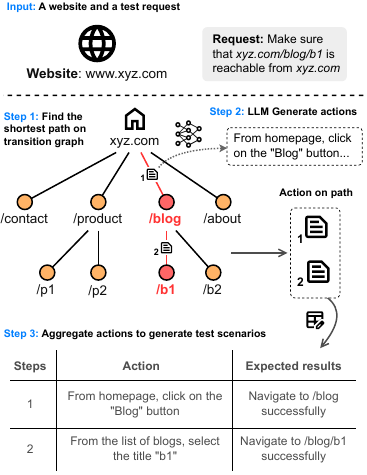}
    \caption{Navigation test scenario generation}
    \label{fig:navigate_test_gen}
\end{figure}

Given a website, the process begins with the generation of a \textit{screen transition graph} through Selenium-based web crawling. Figure \ref{fig:navigate_test_gen} illustrates the process of generating navigation test scenarios. This process takes two inputs: a website and a test request. The test request specifies that a particular webpage (the destination) must be reachable from another specific webpage (the start). The generation process consists of three key steps: \textbf{Path Identification}, \textbf{Action Generation}, and \textbf{Scenario Construction}.

\textbf{Path Identification.} 
In this step, the shortest path from the start node to the destination node in the transition graph is identified. To achieve this, we employ Dijkstra's algorithm \cite{dijkstra1959}, which is well-suited for finding the shortest path in a graph with non-negative edge weights. This ensures an efficient and optimal navigation route between the specified web pages.

\textbf{Action Generation.} 
For each edge along the identified path, a Large Language Model (LLM) is utilized to generate detailed descriptions of the interactions required to navigate from one webpage to another. These descriptions include user actions such as clicking buttons or following links, ensuring a clear and actionable sequence of steps. The use of LLMs enhances the accuracy and comprehensiveness of the generated actions.

\textbf{Scenario Construction.}
The actions generated for each edge are aggregated to construct a comprehensive test scenario. This scenario comprises multiple steps, each describing a specific interaction and its expected outcome. The result is a detailed and executable test case that validates the navigability between the start and destination web pages. This structured approach ensures that the test scenarios are both thorough and practical for real-world application.

\subsection{Form interaction test generation}

\begin{figure}[ht]
    \centering
    \includegraphics[width=1\linewidth]{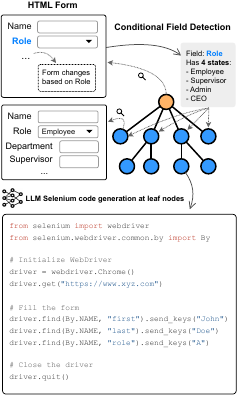}
    \caption{Selenium test case generation via state graph.}
    \label{fig:reaction_form}
\end{figure}

\textbf{Form Types.} 
We categorize forms into two types: \textit{simple forms} and \textit{dynamic forms}. A \textit{simple form} consists solely of input fields and maintains a static structure, unaffected by user interactions. In contrast, a \textit{dynamic form} includes conditional fields that, when interacted with, can alter the form's content or structure. For instance, consider an employee information form (Figure \ref{fig:reaction_form}): selecting "Supervisor" as the employee's role may reveal additional fields specific to supervisors, while selecting "CEO" might display a different set of fields. Consequently, dynamic forms can exhibit multiple states, each representing a unique configuration of the form's fields. Identifying these conditional fields and exploring all possible states of the form are essential for generating comprehensive test cases. Notably, a simple form can be viewed as a special case of a dynamic form with only one state. This section details the process of identifying conditional fields, exploring the states of dynamic forms, and generating Selenium code to test all possible states of the form.


\begin{algorithm}[ht]
\caption{Explore Form Interactions}
\label{alg:form_exploration}
\begin{algorithmic}[1]
\REQUIRE Root form $\mathcal{F}_{root}$
\STATE $\mathcal{E} \gets$ \texttt{FindElements}($\mathcal{F}_{root}$)
\FORALL{$e \in \mathcal{E}$}
    \STATE $\mathcal{H}_{old} \gets$ \texttt{GetHTML}($\mathcal{F}_{root}$)
    \STATE \texttt{InteractWithElement}($e$)
    \STATE $\mathcal{H}_{new} \gets$ \texttt{GetHTML}($\mathcal{F}_{root}$)
    \IF{$\mathcal{H}_{new} \neq \mathcal{H}_{old}$}
        \STATE \texttt{SaveFormState}($\mathcal{F}_{root}, e$)
        \STATE \texttt{ExploreFormInteractions}($\mathcal{F}_{root}$) \COMMENT{Recursive call}
    \ENDIF
\ENDFOR
\end{algorithmic}
\end{algorithm}

\textbf{Form State Exploration.} 
Figure \ref{fig:reaction_form} illustrates the process of form-filling test generation, and Algorithm \ref{alg:form_exploration} details the steps for conditional field detection and state exploration. The algorithm begins by identifying all interactable elements within the root form. For each element, it captures the current state of the form's HTML, simulates an interaction with the element, and then captures the updated HTML. If the HTML changes after the interaction, the algorithm saves the new state of the form and recursively explores further interactions within the updated form. This process continues until all possible interactions and resulting states of the form are exhaustively explored. By systematically identifying conditional fields and generating a comprehensive set of form states, the algorithm enables the creation of Selenium test cases for all possible configurations of the form.

\textbf{Recursive Call in States.} 
Algorithm \ref{alg:form_exploration} employs a recursive call to explore form states, which necessitates backtracking to the previous state upon exiting the recursion. However, backtracking is not directly supported when interacting with the Selenium web driver. To address this limitation, we maintain a list of interactions that lead to the current state. Upon exiting the recursive call, we use this list to replay the sequence of interactions required to return to the state preceding the recursive call. From this point, the algorithm continues to explore other states of the form, ensuring comprehensive coverage of all possible configurations.



\section{Form-filling Benchmark Creation}
\label{sec:data_creation}



\begin{figure*}[ht]
    \centering
    \includegraphics[width=1\linewidth]{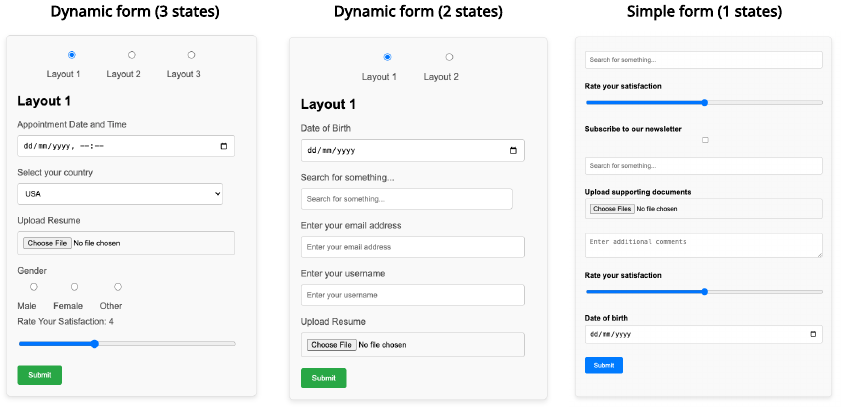}
    \caption{Examples of the two types of synthetic forms in the benchmark: dynamic forms and simple forms.}
    \label{fig:dynamic_form_example}
\end{figure*}

\begin{table*}[ht!]
    \centering
    \caption{Evaluation Criteria for Site-navigation Test Cases.}
    \label{tab:evaluation_criteria}
    \begin{tabular}{ll}
        \toprule
        \textbf{Criteria} & \textbf{Description} \\
        \midrule
        \textbf{Completeness} & Ensures all necessary navigation steps are covered, preventing missing transitions. \\
        \textbf{Accuracy of Expected Results} & Verifies that expected outcomes are clearly defined and align with intended functionality. \\
        \textbf{User Experience} & Assesses whether navigation is intuitive, efficient, and user-friendly. \\
        \textbf{Robustness} & Evaluates handling of unexpected inputs, errors, and edge cases. \\
        \textbf{Clarity and Organization} & Checks if steps are well-structured, logically ordered, and easy to follow. \\
        \bottomrule
    \end{tabular}
    
\end{table*}

\subsection{Overview}
\begin{figure*}[ht]
    \centering
    \includegraphics[width=1\linewidth]{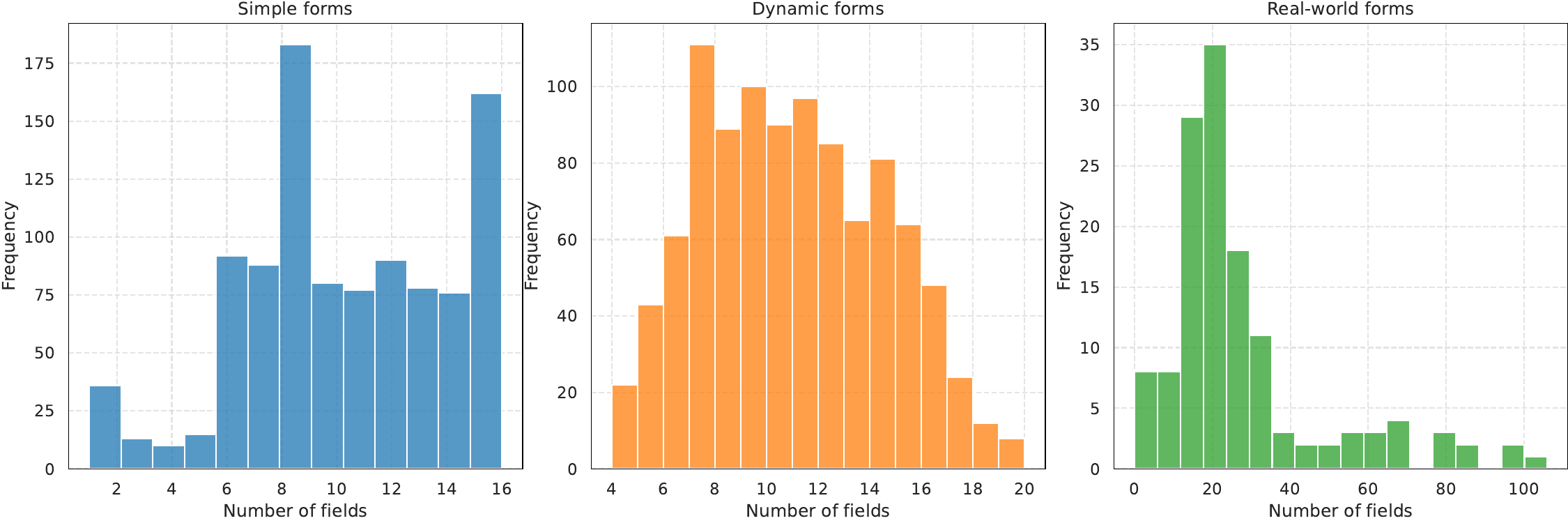}
    \caption{Distribution of the number of fields across different form types in our benchmark.}
    \label{fig:num_field_distribute}
\end{figure*}

The dataset used in the experiments consists of two distinct parts, enabling a comprehensive evaluation of the system’s ability to handle diverse forms and interaction scenarios. Specifically, we developed a dataset to assess the performance of form-interaction test generation approaches. The dataset comprises two main components: (1) rule-based synthetic forms and (2) real-world forms. Figure \ref{fig:num_field_distribute} shows the distribution of the number of fields across different form types in our benchmark.

\subsection{Rule-based Synthetic Forms.}
\begin{figure}[ht]
    \centering
    \includegraphics[width=1\linewidth]{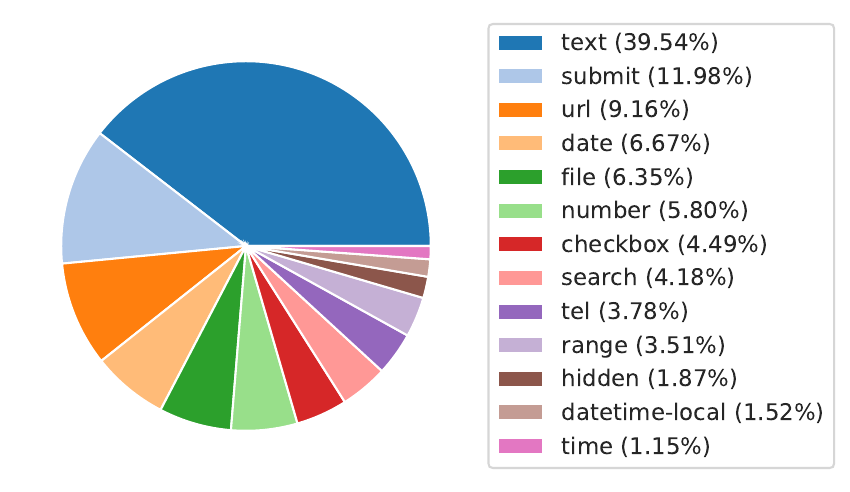}
    \caption{Distribution of field types in our synthetic forms.}
    
    \label{fig:field_type_distribute}
    
\end{figure}

The first part of the dataset consists of 2,000 HTML forms generated using rule-based methods (1,000 simple forms and 1,000 dynamic forms). These forms were designed to encompass a wide range of structures and complexities, ensuring a robust evaluation of the system. The dataset includes both simple forms, such as basic login and registration forms, and dynamic forms, which dynamically change based on user interactions. For each form type, the "required" property was considered in two distinct ways. In the direct approach, the "required" attribute was explicitly defined within the \texttt{input} tags. In the indirect approach, this property was indicated through other HTML elements, such as \texttt{span} elements, which are often used to visually signal required fields. The diversity in structures and constraints ensures that the dataset captures a broad spectrum of real-world scenarios, providing a solid foundation for evaluating the system’s performance. To generate these forms, input tags were selected from a predefined \textit{input pool} containing 200 elements. To enhance the realism of the generated forms, we applied weighted random sampling based on the prevalence of input elements in real-world websites. Specifically, input elements were categorized into six groups according to their frequency of occurrence: \textit{essential}, \textit{very common}, \textit{common}, \textit{moderately common}, \textit{less common}, and \textit{rare}. For instance, \texttt{text} input fields, commonly used for collecting emails or usernames, are assigned higher weights compared to less frequently used inputs such as \texttt{month} or \texttt{week}. For simple forms, input elements were selected randomly while adhering to the predefined weights. In contrast, dynamic forms were constructed by grouping inputs into 2 to 4 sub-forms, with only one sub-form initially visible. The hidden sub-forms are displayed dynamically in response to specific user actions, achieved by modifying the \texttt{display} properties of the HTML elements. An example of our rule-based synthetic forms is illustrated in Figure~\ref{fig:dynamic_form_example}. The distribution of the field types in our synthetic forms is shown in Figure \ref{fig:field_type_distribute}.

\subsection{Real-World Forms.}

To complement the synthetic dataset, we collected a total of 133 forms from real-world websites. These forms were sourced from a diverse range of domains, including e-commerce platforms, content management systems, and social networking websites. By incorporating real-world variability and structural complexity, this dataset serves as a realistic benchmark for evaluating the system’s performance in practical scenarios. The inclusion of real-world forms ensures that the proposed approach is tested against naturally occurring challenges such as varying form layouts, dynamic content loading, and diverse validation mechanisms. Together with the synthetic dataset, these real-world forms provide a comprehensive testbed for assessing the system’s effectiveness across a wide range of web testing challenges.


\section{Experiments}

\subsection{Dataset}

For \textbf{site-navigation} test generation, we curated a set of 10 real-world websites with diverse and complex sitemaps suitable for our experiments. For each website, the homepage was selected as the starting point, and destination pages were randomly sampled. The experiment involved using an LLM to analyze the screen transition graph and generate test case scenarios to ensure that the destination pages were reachable from the starting point. The quality of these generated test scenarios was then evaluated using the commercial LLM GPT-4o. For \textbf{form-filling} test generation, we conducted experiments on our curated benchmark dataset, as described in Section~\ref{sec:data_creation}. 

\subsection{Models}

We evaluated both closed-source commercial LLMs and open-source LLMs.  
For \textbf{closed-source LLMs}, we utilized GPT-4o~\cite{openai2024gpt4ocard} for generating test cases for both site navigation and form filling.  
For \textbf{open-source LLMs}, we employed DeepSeek Distill (Qwen-7B, LLaMA3-8B/70B)~\cite{deepseekai2025deepseekr1incentivizingreasoningcapability} for site-navigation test generation, and Qwen2.5-Instruct (7B and 14B)~\cite{bai_qwen_2023} along with Llama3.1-Instruct (8B)~\cite{grattafiori2024llama3herdmodels} for form-filling test generation.


\subsection{Metrics}

For \textbf{site-navigation} test scenario generation, we followed prior work~\cite{gu2025surveyllmasajudge} and employed GPT-4o to evaluate the quality of the generated test scenarios. The evaluation was based on five predefined criteria, as detailed in Table~\ref{tab:evaluation_criteria}.

For \textbf{form-filling} test case generation, we assessed performance using two key metrics:

\begin{itemize}
    \item \textbf{Pass Accuracy}: The proportion of successful interactions out of the total interactions performed. We report two variants:  
    \begin{itemize}
        \item \textbf{Micro-accuracy}: Computed as the total number of successful interactions across all test cases divided by the total number of interactions.  
        \item \textbf{Macro-accuracy}: Computed as the average accuracy across individual test cases, where each test case's accuracy is calculated independently.  
    \end{itemize}
    \item \textbf{Test Coverage}: Defined as the ratio of the number of fields interacted with during testing to the total number of fields in a form, this metric quantifies the proportion of the form's functionality that is validated by the test cases.
\end{itemize}

\section{Results}

\begin{table*}[ht]
    \centering
    \begin{tabular}{clccc}
    \toprule
         Model&   Setting&Micro Accuracy&Macro Accuracy &Coverage\\
         \midrule
         \multirow{3}{*}{GPT4o} &   Simple form&\textbf{95.17}& \textbf{94.74}&\textbf{69.49}\%\\
         &   Dynamic form&\textbf{79.96}& \textbf{78.86}&56.13\%\\
         &   Real-world form&\textbf{91.37}& \textbf{90.19}& \textbf{37.12}\%\\
    \midrule
 \multirow{3}{*}{Qwen2.5-Instruct (7B)}& Simple form& 84.79& 83.69 & 19.68\%\\
 & Dynamic form& 54.22& 55.82& 8.13\%\\
 & Real-world form& 83.95& 79.29&9.45\%\\
\midrule
 \multirow{3}{*}{Llama3.1-Instruct (7B)}& Simple form& 30.11& 30.17 & 67.63\%\\
 & Dynamic form&  42.79 & 45.43 &  \textbf{68.45}\%\\
 & Real-world form& 31.66 & 29.46 & 28.25\%\\
 \bottomrule
    \end{tabular}
    \caption{Performance of Form-filling test generation on synthetic forms (simple and dynamic) and real-world form.}
    \label{tab:acc_result}
\end{table*}

\begin{figure*}[ht]
    \centering
    \begin{subfigure}[b]{0.32\textwidth}
    \includegraphics[width=1\linewidth]{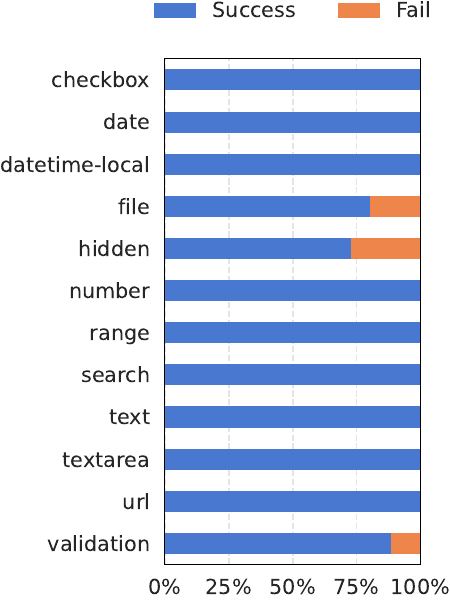}
    \caption{Simple form.}
        \label{fig:subfig1}
    \end{subfigure}
    \hfill 
    \begin{subfigure}[b]{0.32\textwidth}
    \includegraphics[width=1\linewidth]{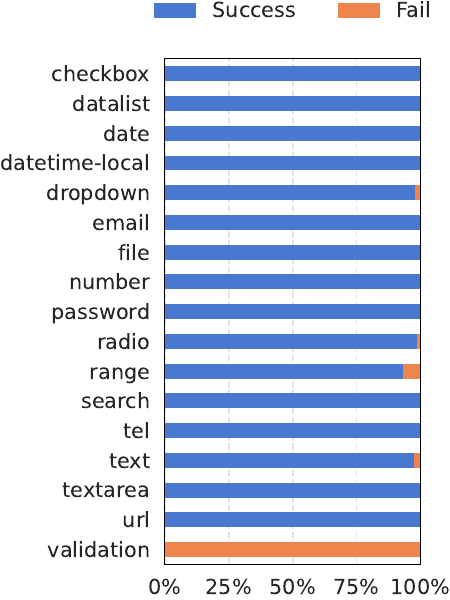}
    \caption{Dynamic form.}
        \label{fig:subfig2}
    \end{subfigure}
    \hfill 
    \begin{subfigure}[b]{0.32\textwidth}
    \includegraphics[width=1\linewidth]{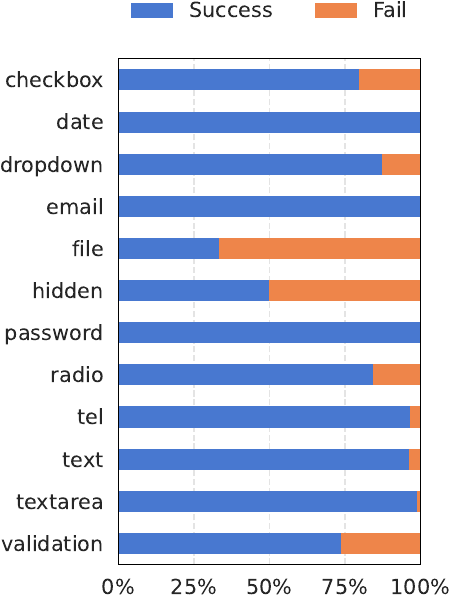}
    \caption{Real-world form.}
        \label{fig:subfig3}
    \end{subfigure}
    \caption{Success/Fail by field type of form interactions using GPT4o.}
    \label{fig:success_fail_by_type}
\end{figure*}

\begin{table}[ht]
\resizebox{\linewidth}{!}{
\begin{tabular}{lllllll}
\toprule
Model & Com & Acc & Exp & Rob& Org & Final \\
\midrule
R1-D(Llama-70B) & 6.78 & 7.83 & 7.12 & 4.78 & 8.67 & 7.01 \\
R1-D(Llama-8B) & 6.04 & 7.37 & 6.44 & 4.17 & 8.08 & 6.41 \\
R1-D(Qwen-7B) & 5.43 & 6.03 & 5.44 & 3.64 & 6.81 & 5.47 \\
GPT4o-mini & \textbf{7.63} & \textbf{8.46} & \textbf{7.61} & \textbf{5.61} & \textbf{8.70} & \textbf{7.60} \\
\bottomrule
\end{tabular}
}
    \caption{Performance of site-navigation test generation evaluated across five criteria: Completeness (Com), Accuracy of Expected Results (Acc), User Experience (Exp), Robustness (Rob), and Clarity and Organization (Org). The evaluated models include variants of Deepseek R1 Distill (R1-D) and GPT-4o-mini.}
    \label{tab:result_navigation}

\end{table}

\subsection{Main Result}

\textbf{Site-navigation Test.} Table \ref{tab:result_navigation} presents the evaluation results of different models in generating site-navigation test scenarios across five criteria. GPT-4o-mini achieved the highest overall performance, excelling in clarity, accuracy, and completeness. R1-D (Llama-70B) followed closely, with strong clarity and accuracy but slightly lower robustness. R1-D (Llama-8B) performed moderately, while R1-D (Qwen-7B) had the lowest scores, particularly in accuracy and robustness. The results indicate that larger models generally perform better, producing more structured and accurate test cases. However, robustness remains a challenge across all models, suggesting room for improvement in handling unexpected navigation issues.

\textbf{Form-filling Test.} Table \ref{tab:acc_result} presents the performance of different models in generating form-filling test cases across three settings: simple forms, dynamic forms, and real-world forms. GPT-4o achieved the highest accuracy across all settings, particularly excelling in simple and real-world forms, demonstrating its ability to handle both structured and complex scenarios effectively. Qwen2.5-Instruct (7B) performed reasonably well on simple and real-world forms but struggled with dynamic forms, showing a significant drop in accuracy and coverage. Llama3.1-Instruct (7B) had the lowest overall performance, particularly in macro accuracy, though it maintained relatively high coverage in simple and dynamic forms. Overall, the results suggest that larger and more advanced models generalize better across different form types, with GPT-4o significantly outperforming smaller models. However, most models show a decline in performance on dynamic forms, highlighting challenges in handling complex interactions in web forms.

\subsection{Analysis} 

Figure \ref{fig:success_fail_by_type} shows the GPT4o's success and failure rates of different form field types across simple, dynamic, and real-world forms. The results indicate that while the method effectively handles most common form fields, certain field types exhibit higher error rates. Notably, the \textit{hidden} and \textit{file} fields show consistent failures across all settings, with real-world forms experiencing the highest failure rates. Additionally, validation errors are more prominent, particularly in dynamic and real-world forms, suggesting challenges in handling complex form validation rules. Despite these issues, the method demonstrates strong performance for frequently used field types such as \textit{text}, \textit{checkbox}, \textit{date}, and \textit{number}, indicating its reliability for most practical applications. Further refinements in handling edge cases, particularly for specialized field types and validation mechanisms, could enhance overall robustness.



\section{Conclusion}

This paper presented an automated system for generating test cases for site navigation and form filling, addressing key challenges in web application testing. By leveraging screen transition graphs and state graphs combined with large language models (LLMs), the system provides a scalable and efficient solution for modeling navigation flows and handling complex form interactions. Experimental results demonstrated the system's effectiveness in improving test coverage and robustness, supported by a comprehensive dataset for evaluation. The integration of both commercial and open-source LLMs highlighted their potential in automating test case generation while maintaining high accuracy. This work advances the field of web application testing by offering a unified framework that bridges navigation modeling and form validation, paving the way for more intelligent and adaptive testing solutions in the future.

\bibliographystyle{IEEEtran} 
\bibliography{references} 

@article{marchetto2012,
  author = {Marchetto, A. and Tonella, P. and Ricca, F.},
  title = {A Survey on Web Application Testing},
  journal = {ACM Computing Surveys},
  volume = {44},
  number = {3},
  pages = {1--36},
  year = {2012}
}

@inproceedings{ricca2001,
  author = {Ricca, F. and Tonella, P.},
  title = {Analysis and Testing of Web Applications},
  booktitle = {Proceedings of the International Conference on Software Engineering},
  volume = {25},
  number = {3},
  pages = {25--34},
  year = {2001}
}

@article{andrews2005,
  author = {Andrews, A. and Offutt, J. and Alexander, R.},
  title = {Test Generation for Web Applications},
  journal = {IEEE Transactions on Software Engineering},
  volume = {31},
  number = {3},
  pages = {187--202},
  year = {2005}
}

@article{smith2022,
  author = {Smith, J. and Taylor, R.},
  title = {Automated Frameworks for Dynamic Web Testing},
  journal = {Software Testing Journal},
  volume = {37},
  number = {1},
  pages = {45--67},
  year = {2022}
}

@inproceedings{lee2022,
  author = {Lee, K. and Johnson, S.},
  title = {Leveraging Generative AI for Automated Test Case Creation},
  booktitle = {Proceedings of ICSE},
  year = {2022},
  pages = {198--207}
}

@article{dijkstra1959,
author = {Dijkstra, E. W.},
title = {A note on two problems in connexion with graphs},
year = {1959},
issue_date = {December  1959},
publisher = {Springer-Verlag},
address = {Berlin, Heidelberg},
volume = {1},
number = {1},
issn = {0029-599X},
url = {https://doi.org/10.1007/BF01386390},
doi = {10.1007/BF01386390},
journal = {Numer. Math.},
month = dec,
pages = {269–271},
numpages = {3}
}

@misc{openai2024gpt4ocard,
      title={GPT-4o System Card}, 
      author={OpenAI},
      year={2024},
      eprint={2410.21276},
      archivePrefix={arXiv},
      primaryClass={cs.CL},
      url={https://arxiv.org/abs/2410.21276}, 
}

@misc{grattafiori2024llama3herdmodels,
      title={The Llama 3 Herd of Models}, 
      author={Aaron Grattafiori et al.},
      year={2024},
      eprint={2407.21783},
      archivePrefix={arXiv},
      primaryClass={cs.AI},
      url={https://arxiv.org/abs/2407.21783}, 
}

@inproceedings{he-etal-2024-webvoyager,
    title = "{W}eb{V}oyager: Building an End-to-End Web Agent with Large Multimodal Models",
    author = "He, Hongliang  and
      Yao, Wenlin  and
      Ma, Kaixin  and
      Yu, Wenhao  and
      Dai, Yong  and
      Zhang, Hongming  and
      Lan, Zhenzhong  and
      Yu, Dong",
    editor = "Ku, Lun-Wei  and
      Martins, Andre  and
      Srikumar, Vivek",
    booktitle = "Proceedings of the 62nd Annual Meeting of the Association for Computational Linguistics (Volume 1: Long Papers)",
    month = aug,
    year = "2024",
    address = "Bangkok, Thailand",
    publisher = "Association for Computational Linguistics",
    url = "https://aclanthology.org/2024.acl-long.371/",
    doi = "10.18653/v1/2024.acl-long.371",
    pages = "6864--6890",
}

@article{openai2023gpt4,
  author    = {OpenAI},
  title     = {GPT-4 Technical Report},
  year      = {2023},
  journal   = {arXiv preprint},
  archivePrefix = {arXiv},
  eprint    = {2303.08774}
}

@misc{autogpt2022,
  author    = {AutoGPT},
  title     = {AutoGPT},
  year      = {2022}
}

@article{qin2023toollm,
  author    = {Yujia Qin and Shihao Liang and Yining Ye and Kunlun Zhu and Lan Yan and Yaxi Lu and Yankai Lin and Xin Cong and Xiangru Tang and Bill Qian, et al.},
  title     = {ToolLLM: Facilitating Large Language Models to Master 16000+ Real-World APIs},
  journal   = {arXiv preprint},
  year      = {2023},
  archivePrefix = {arXiv},
  eprint    = {2307.16789}
}

@article{schick2023toolformer,
  author    = {Timo Schick and Jane Dwivedi-Yu and Roberto Dessì and Roberta Raileanu and Maria Lomeli and Luke Zettlemoyer and Nicola Cancedda and Thomas Scialom},
  title     = {Toolformer: Language Models Can Teach Themselves to Use Tools},
  journal   = {arXiv preprint},
  year      = {2023},
  archivePrefix = {arXiv},
  eprint    = {2302.04761}
}

@article{nakano2021webgpt,
  author    = {Reiichiro Nakano and Jacob Hilton and Suchir Balaji and Jeff Wu and Long Ouyang and Christina Kim and Christopher Hesse and Shantanu Jain and Vineet Kosaraju and William Saunders, et al.},
  title     = {WebGPT: Browser-Assisted Question-Answering with Human Feedback},
  journal   = {arXiv preprint},
  year      = {2021},
  archivePrefix = {arXiv},
  eprint    = {2112.09332}
}

@article{gur2023realworld,
  author    = {Izzeddin Gur and Hiroki Furuta and Austin Huang and Mustafa Safdari and Yutaka Matsuo and Douglas Eck and Aleksandra Faust},
  title     = {A Real-World WebAgent with Planning, Long Context Understanding, and Program Synthesis},
  journal   = {arXiv preprint},
  year      = {2023},
  archivePrefix = {arXiv},
  eprint    = {2307.12856}
}

@article{zhou2023webarena,
  author    = {Shuyan Zhou and Frank F Xu and Hao Zhu and Xuhui Zhou and Robert Lo and Abishek Sridhar and Xianyi Cheng and Yonatan Bisk and Daniel Fried and Uri Alon, et al.},
  title     = {WebArena: A Realistic Web Environment for Building Autonomous Agents},
  journal   = {arXiv preprint},
  year      = {2023},
  archivePrefix = {arXiv},
  eprint    = {2307.13854}
}

@article{lu2023chameleon,
  author    = {Pan Lu and Baolin Peng and Hao Cheng and Michel Galley and Kai-Wei Chang and Ying Nian Wu and Song-Chun Zhu and Jianfeng Gao},
  title     = {Chameleon: Plug-and-Play Compositional Reasoning with Large Language Models},
  journal   = {arXiv preprint},
  year      = {2023},
  archivePrefix = {arXiv},
  eprint    = {2304.09842}
}

@article{deng2023mind2web,
  author    = {Xiang Deng and Yu Gu and Boyuan Zheng and Shijie Chen and Samuel Stevens and Boshi Wang and Huan Sun and Yu Su},
  title     = {Mind2Web: Towards a Generalist Agent for the Web},
  journal   = {arXiv preprint},
  year      = {2023},
  archivePrefix = {arXiv},
  eprint    = {2306.06070}
}

@misc{bai_qwen_2023,
    title = {Qwen {Technical} {Report}},
    author = {Bai, Jinze and Bai, Shuai and Chu, Yunfei and Cui, Zeyu and Dang, Kai and Deng, Xiaodong and Fan, Yang and Ge, Wenbin and Han, Yu and Huang, Fei and Hui, Binyuan and Ji, Luo and Li, Mei and Lin, Junyang and Lin, Runji and Liu, Dayiheng and Liu, Gao and Lu, Chengqiang and Lu, Keming and Ma, Jianxin and Men, Rui and Ren, Xingzhang and Ren, Xuancheng and Tan, Chuanqi and Tan, Sinan and Tu, Jianhong and Wang, Peng and Wang, Shijie and Wang, Wei and Wu, Shengguang and Xu, Benfeng and Xu, Jin and Yang, An and Yang, Hao and Yang, Jian and Yang, Shusheng and Yao, Yang and Yu, Bowen and Yuan, Hongyi and Yuan, Zheng and Zhang, Jianwei and Zhang, Xingxuan and Zhang, Yichang and Zhang, Zhenru and Zhou, Chang and Zhou, Jingren and Zhou, Xiaohuan and Zhu, Tianhang},
    year = {2023},
}

@misc{deepseekai2025deepseekr1incentivizingreasoningcapability,
      title={DeepSeek-R1: Incentivizing Reasoning Capability in LLMs via Reinforcement Learning}, 
      author={DeepSeek-AI},
      year={2025},
      eprint={2501.12948},
      archivePrefix={arXiv},
      primaryClass={cs.CL},
      url={https://arxiv.org/abs/2501.12948}, 
}

@misc{gu2025surveyllmasajudge,
      title={A Survey on LLM-as-a-Judge}, 
      author={Jiawei Gu and Xuhui Jiang and Zhichao Shi and Hexiang Tan and Xuehao Zhai and Chengjin Xu and Wei Li and Yinghan Shen and Shengjie Ma and Honghao Liu and Saizhuo Wang and Kun Zhang and Yuanzhuo Wang and Wen Gao and Lionel Ni and Jian Guo},
      year={2025},
      eprint={2411.15594},
      archivePrefix={arXiv},
      primaryClass={cs.CL},
      url={https://arxiv.org/abs/2411.15594}, 
}


\end{document}